\documentclass[preprint,showpacs,preprintnumbers,amsmath,amssymb]{revtex4}

\usepackage{graphicx}
\usepackage{dcolumn}
\usepackage{bm}

\begin{document}


\title{Peripheral interactions of relativistic $^{14}$N nuclei\\ with emulsion nuclei}

\author{T.~V.~Shchedrina}
   \email{shchedrina@lhe.jinr.ru} 
   \affiliation{Joint Insitute for Nuclear Research, Dubna, Russia}
 \author{V.~Bradnova}
   \affiliation{Joint Insitute for Nuclear Research, Dubna, Russia} 
\author{M.~M.~Chernyavsky}
  \affiliation{Lebedev Institute of Physics, Russian Academy of Sciences, Moscow, Russia}  
\author{S.~P.~Kharlamov}
   \affiliation{Lebedev Institute of Physics, Russian Academy of Sciences, Moscow, Russia}
\author{A.~D.~Kovalenko}
   \affiliation{Joint Insitute for Nuclear Research, Dubna, Russia}  
\author{M.~Haiduc}
   \affiliation{Institute of Space Sciences, Magurele, Romania}     
\author{A.~I.~Malakhov}
   \affiliation{Joint Insitute for Nuclear Research, Dubna, Russia} 
\author{G.~I.~Orlova}
   \affiliation{Lebedev Institute of Physics, Russian Academy of Sciences, Moscow, Russia} 
\author{P.~A.~Rukoyatkin}
   \affiliation{Joint Insitute for Nuclear Research, Dubna, Russia} 
\author{V.~V.~Rusakova}
   \affiliation{Joint Insitute for Nuclear Research, Dubna, Russia} 
\author{S.~Vok\'al}
   \affiliation{P. J. \u Saf\u arik University, Ko\u sice, Slovak Republic}
\author{A.~Vok\'alov\'a}
   \affiliation{P. J. \u Saf\u arik University, Ko\u sice, Slovak Republic}    
 \author{P.~I.~Zarubin}
     \email{zarubin@lhe.jinr.ru}    
     \homepage{http://becquerel.lhe.jinr.ru}
   \affiliation{Joint Insitute for Nuclear Research, Dubna, Russia} 
 \author{I.~G.~Zarubina}
   \affiliation{Joint Insitute for Nuclear Research, Dubna, Russia}   

\date{\today}

\begin{abstract}
\indent  The results of investigations of the dissociation of a $^{14}$N nucleus of
 momentum 2.86~A~GeV/c  in photo-emulsion  are presented. The main characteristics of
 these reactions, that is  the cross sections  for various fragmentation channels, are given.
 The fragmentation was analyzed  by means of an invariant approach. The momentum and
 correlation characteristics of $\alpha$ particles for the $^{14}$N$\rightarrow$3$\alpha$+X
 channel in the laboratory system and the rest systems of 3$\alpha$ particles were considered.
 The results obtained for the $^{14}$N nucleus are compared  with similar data for the $^{12}$C
 and $^{16}$O nuclei.\par
\end{abstract}

\maketitle
\section{\label{sec:level1}Introduction}
\indent The fragmentation of relativistic nuclei is a source of information about their
 structure. The nuclear photo-emulsion method  makes it possible to study in detail the
 fragmentation of a projectile thanks to a high resolution capability of emulsion and the
 detection of secondaries  in a 4$\pi$ geometry. The registration of all charged particles
 and their identification enable one to explore the isotopic composition of fragments and the
 projectile fragmentation channels. In the present paper, we give a detailed examination both
 of the decays conventionally called \lq\lq white\rq\rq starts because they have no
 target-nucleus fragments and of the decays involving the production of  several
 target-nucleus fragments. The main criterion for selecting the appropriate  events used
 for the study of the projectile fragmentation is the requirement of conservation of the
 primary charge in a narrow fragmentation cone and the absence of produced charged particles.\par
 \section{\label{sec:level2}Experiment}
\indent  A stack of layers  of BR emulsion of a relativistic sensitivity was exposed to a
 beam of $^{14}$N nuclei accelerated to a momentum of 2.86~A~GeV/c  at the Nuclotron of the
 Laboratory of High Energy Physics (JINR). The layer thickness and dimensions were 600$\mu$m and
10$\times$20 cm$^2$, respectively. The exposed beam was pointed parallel to the emulsion
 plane along its long side. Events were sought by viewing over the track length  which
 provided the accumulation of statistics without selection. This made it possible to
 determine the mean free path of $^{14}$N interactions in emulsion. The fragment emission
 angles were measured by the microscope MPE-11 (Moscow, FIAN). Coordinate sensors were set
 on the X,Y, and Z axes to transmit information  to a personal computer which provided its
 processing. The Z=1 projectile fragments  were separated from the Z=2 fragments by sight
 because a single ionization of relativistic single-charged particles is reliably
 discriminated from a four-fold ionization of particles with charge 2. Fragments with
 Z=3,...,7 were separated by counting $\delta$ electrons. The momenta of single- and
 two-charged fragments with emission angles up to 4$\times$ were determined from
 multiple coulomb scattering measurements. The results of these measurements were utilized
 to identify the hydrogen and helium isotopes involved in the $^{14}$N fragment composition.\par
 
\section{\label{sec:level3}The mean free path of interactions}
 
\indent On the viewed track length of 123.21 m we found 960 events of interactions of the
 $^{14}$N nucleus with the emulsion nuclei. Thus the mean free path of a $^{14}$N in emulsion
 is $\lambda_{N}=(13\pm 0.4)$cm. This value and the $\lambda_{A}$ values for a number of
 other projectiles in emulsion obtained earlier in \cite{Adamovich99},\cite{AdamovichPh04}
 are displayed in Table \ref{tab:1} in which there are also the $\lambda_{A}$ values  calculated by the
 Bradt-Peters equation \cite{Bradt}.\par
\begin{table}
\caption{\label{tab:1}The mean free path of $\lambda$ in nuclear emulsion}

\begin{tabular}{l|c|c|c|c}
\hline\noalign{\smallskip}
\hline\noalign{\smallskip}
~$Nuclei$~&~p$_{0}$,~A~GeV/c~&~$\lambda_{theor}$, cm.~&~$\lambda_{exp}$, cm.~&~Ref.~  \\
\hline\noalign{\smallskip}
~$^4$He~& 4.5 & 19.6 & 19.5$\pm$0.3 & \cite{Adamovich99}  \\
~$^6$Li~& 4.5 & 16.5 & 14.1$\pm$0.4 & \cite{Adamovich99}   \\
~$^7$Li~& 3.0 & 15.9 & 14.3$\pm$0.4 &  \cite{AdamovichPh04}  \\
~$^{12}$C~& 4.5 & 13.5 & 13.7$\pm$0.5 & \cite{Adamovich99}  \\
~$^{14}$N~& 2.9 & 13.0 & 13.0$\pm$0.4 &~present~paper~   \\
~$^{16}$O~& 4.5 & 12.1 & 13.0$\pm$0.5 & \cite{Adamovich99}   \\
~$^{22}$Ne~& 4.1 & 10.6 & 10.2$\pm$0.1 & \cite{Adamovich99}  \\
~$^{24}$Mg~& 4.5 & 10.0 & 9.6$\pm$0.4 & \cite{Adamovich99}   \\
\hline\noalign{\smallskip}
\hline\noalign{\smallskip}
\end{tabular}
\end{table}

\section{\label{sec:level4}Topology in the relativistic $^{14}$N fragmentation}

\indent Of 950 interactions found, we selected events in which the total fragment
 charge was equal to the Z$_{0}$=7 fragment charge and there were no produced particles.
 The selected events are divided in two classes. The events of the type of
 \lq\lq white\rq\rq star (44 events) and the interactions involving the production
 of one or a few target-nucleus fragments (61 events) belong to the first class.
 The name \lq\lq white\rq\rq is a conventional one that refers to the interactions
 in which there are neither target fragments nor produced particles. They are
 produced in case when the energy transferred to a fragmenting nucleus is minimal
 which leads to a distraction of inter-cluster bounds but, as a rule, the bounds
 inside the clusters remain unaffected. This is the reason  for which they are of
 special interest for the study of the cluster structure of nuclei. The other
 class of events, we are interested in, reveals a simultaneous breakup of both
  interacting nuclei which results in the production of target-nucleus fragments.
 The projectile fragments are emitted mainly within a narrow forward cone the
 opening angle of which is described by the relation (\ref{eq1})
\begin{equation}
 { sin\Theta_{f}=\frac{0.2~GeV/c}{p_{0}}\label{eq1}} 
 \end{equation}
where 0.2~GeV/c is the value of the Fermi momentum and p$_{0}$ the momentum per
 nucleon of the projectile. In our case, at p$_{0}$=2.86~A~GeV/c the fragmentation
 angle is 4$^{\circ}$.\par
 \indent Table \ref{tab:2} shows the charge multi-fragmentation topology which was
 studied for the events satisfying the above-mentioned conditions. The upper line
 is the Z$>$2 fragment charge, the second line is  the number of single-charged
 fragments , the third one the number of two-charged fragments, and the fourth and
 fifth lines are the number of the detected events with a given topology for
 \lq\lq white\rq\rq stars and events with target-nucleus excitation for each
 channel, respectively. The two last lines present the total number of interactions
 calculated in absolute values and in percent.\par 
\begin{table}
\caption{\label{tab:2}The charge topology distribution of the
\lq\lq white\rq\rq stars and the interactions involving the target-nucleus
 fragment production in the $^{14}$N dissociation at 2.86~A~GeV/c momentum.}
\begin{tabular}{ l@{\hspace{5mm}}|@{\hspace{2mm}}c@{\hspace{2mm}}|@{\hspace{2mm}}c@{\hspace{2mm}}|
@{\hspace{2mm}}c@{\hspace{2mm}}|@{\hspace{2mm}}c@{\hspace{2mm}}|
@{\hspace{2mm}}c@{\hspace{2mm}}|
@{\hspace{2mm}}c@{\hspace{2mm}}|
@{\hspace{2mm}}c@{\hspace{2mm}}|@{\hspace{2mm}}c@{\hspace{2mm}}}
\hline\noalign{\smallskip}
\hline\noalign{\smallskip}
~Z$_{fr}$~& 6 & 5 & 5 & 4 & 3 & 3 & -- & -- \\
~N$_{Z=1}$~& 1 & -- & 2 & 1 & 4 & 2 & 3 & 1   \\
~N$_{Z=2}$~& -- & 1 & -- &  1 & -- & 1 & 2 & 3 \\
~N$_{W.S.}$~& 13 & 4 & 3 & 1 & 1 & 1 & 6 & 15  \\
~N$_{t.f.}$~& 15 & 1 & 3 & 3 & -- & 2 & 5 & 32  \\
~N$_{\sum}$~& 28 & 5 & 6 & 4 & 1 & 3 & 11 & 49   \\
~N$_{\sum, \%}$~& 26 & 5 & 5 & 4 & 1 & 3 & 10 & 46  \\
\hline\noalign{\smallskip}
\hline\noalign{\smallskip}
\end{tabular}
\end{table}
\indent The analysis of the data of Table \ref{tab:2} shows that the number of
 channels involving Z$>$3 fragments for the \lq\lq white\rq\rq stars is larger by
 about a factor of 1.5 than that for the events accompanied by a target breakup.
 On the contrary, for the 2+2+2+1 charge configuration channel this number is
 smaller by about a factor of 1.5. Thus, in the events with target breakup, the
 projectile fragments more strongly  than in the \lq\lq white\rq\rq stars. The
 data of Table \ref{tab:2}  points to the predominance of the channel with the
 2+2+2+1 charge configuration (49 events) which has been studied in more detail.
 The mean free path for this channel $\lambda_{3He+H}$($^{14}$N)=2.5$\pm$0.36 m.
 An analogous value for the carbon nucleus is larger by a factor of 4,
 $\lambda_{3He}$=10.3$\pm$1.9 m. The obtained results show that the $^{14}$N
 nucleus constitutes a very effective source for the production of 3$\alpha$ system.\par
\begin{figure*}
\footnotesize
 \centerline{\begin{tabular}{cc}
 \includegraphics[height=50mm]{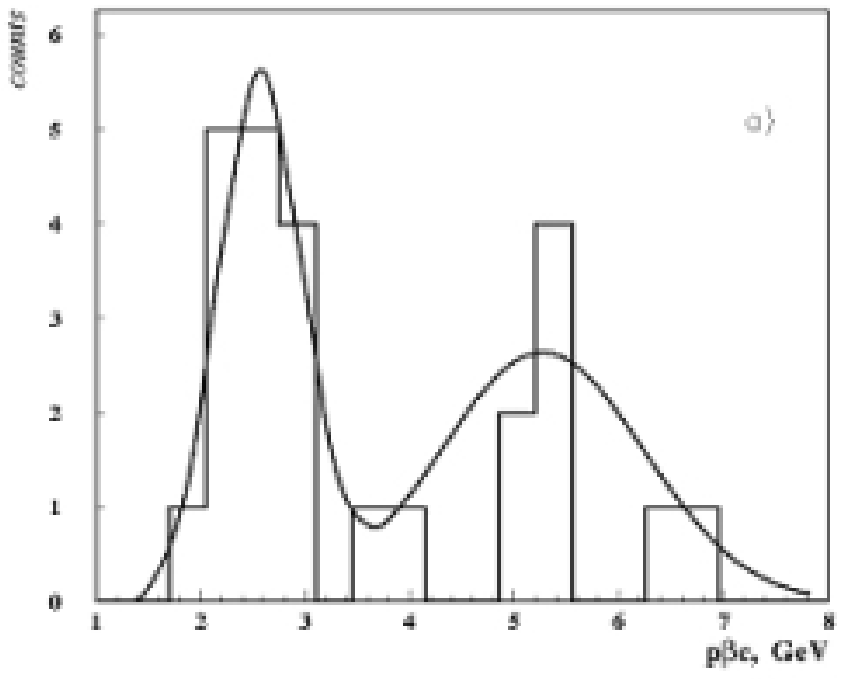} &
 \includegraphics[height=50mm]{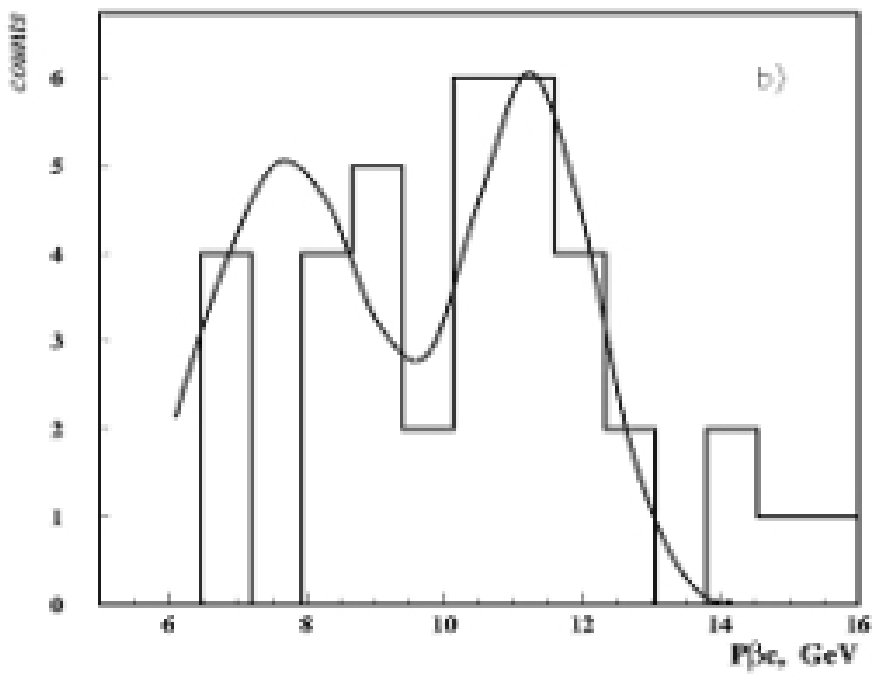}\\
 \end{tabular}}
\caption{\label{fig:1}Momentum separation of single- a) and double- b) charged
 fragments according to p$\beta$c measurements for the $^{14}$N nucleus of a momentum of
 2.86~A~GeV/c. The continuous line is a Gaussian description  by the least square
 method.}
\end{figure*}

\indent The hydrogen and helium isotopes were separated basing on the results of
 measurements of their momenta (p$\beta$c) under the assumption that the spectator
 target-nucleus fragments conserve their momentum per nucleon which is equal to the
 primary one, that is,  A$_{fr}$=p$\beta$c$_{exp}$/p$_{0}\beta$c. The multiple
 coulomb scattering method used for the determination of the momenta is based on
 the suggestion that the rms of a  $<|D|>$ over a cell length of t is associated
 with the p$\beta$c value by the equation (\ref{eq2})
 \begin{equation}
 { <|D|>=\frac{Z_{f}\cdot K\cdot t^{\frac{3}{2}}}{573\cdot p\beta c}\label{eq2}} 
 \end{equation}
where Z is the charge, p the momentum, $\beta$c the particle velocity, and the
 scattering constant  the value of which is known. Distortions and a false
 scattering were taken into account  by a so-called $\rho$ method \cite{web}.
  In multiple coulomb scattering the p$\beta$c distribution for individual
 particles with the same charge and momentum must be close to the normal one.
 Therefore the p$\beta$c distribution for a group of fragments with identical
 velocity and charge, but with different masses, must be a superposition  of
 several normal distributions. Fig. \ref{fig:1}(a,b) presents the results of
 measurements of a multiple-particle scattering for single- and doubly-charged
 fragments, respectively. The measured momenta for single-charged fragments are
 well approximated by the sum of two Gaussians the peaks of which are located at
 2.6~GeV/c and 5.6~GeV/c  and correspond to $^1$H and $^2$H isotopes
 Fig. \ref{fig:1}(a). The ratio of the $^1$H and $^2$H isotope yields thus
 obtained is nearly 2$:$1. This points out that, in our case, the deuteron fraction
 is seen to be noticeably smaller as compared to the cases of  the relativistic
 $^6$Li (2+1 channel) and $^{10}$B (2+2+1 channel ) fragmentation where the proton
 and deuteron yields are about  the same.\par
\indent Fig. \ref{fig:1}(b) gives the measured p$\beta$c distribution for 37
 double-charged fragments. This distribution is satisfactorily approximated by
 the sum of two normal distributions which is shown by the continuous line. The
 approximating distribution peaks correspond to the values 7.8 and 11.3 which are
 rather close to the p$\beta$c values which relate to the $^3$He and $^4$He
 isotopes. The fragment yield for $^3$He is about 40\% and for $^4$He - 60\%.
 There are also a few He isotopes (5\% of the total number of interactions) in
 the p$\beta$c range from 14 to 16 MeV  that were identified as $^6$He. The events
 involving the $^6$He production are planned to be analyzed in more detail.\par 
\section{\label{sec:level5}Momentum and correlation characteristics of N$\alpha$
 particle systems} 
 
\begin{figure*}
\footnotesize
 \centerline{\begin{tabular}{cc}
 \includegraphics[height=50mm]{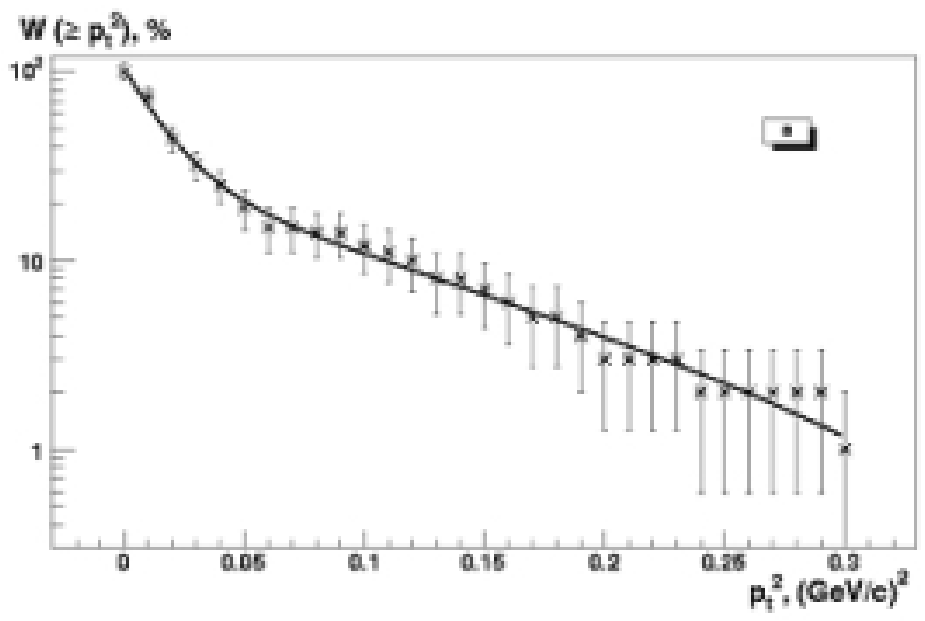} &
 \includegraphics[height=55mm]{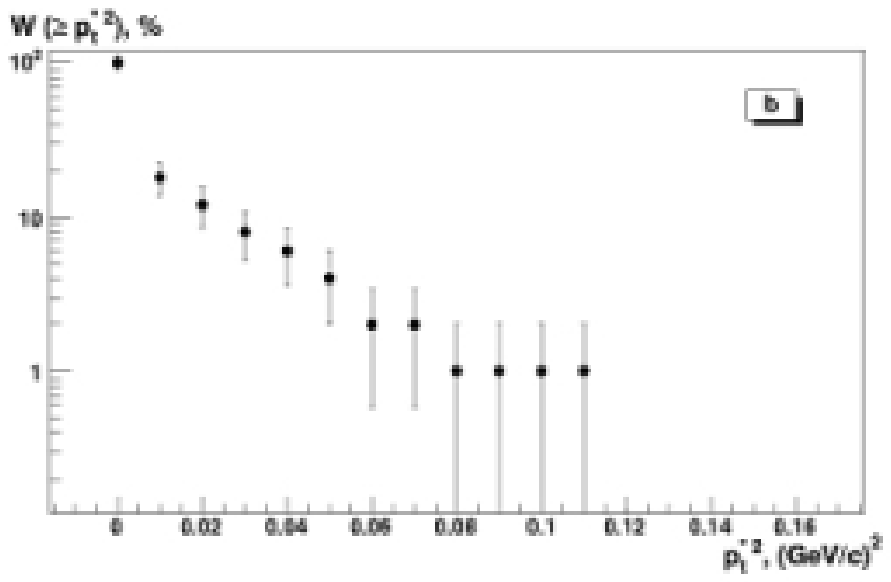}\\
 \end{tabular}}
\caption{\label{fig:2} The P$_{T}^2$ lab.system (a) and P$_{T}^{*2}$ 3$\alpha$
 particle rest system(b) distributions (channel $^{14}$N$\rightarrow$3$\alpha$+X). The continuous line of Fig. \ref{fig:2}a is
 the sum of two Relay distributions.}
\end{figure*}

 \indent We go over to the consideration of the main kinematic characteristics of
 relativistic $\alpha$ particles, the  projectile fragments, from the
 $^{14}$N$\rightarrow$3$\alpha$+X reaction and we compare these characteristics
 with those from the $^{12}$C$\rightarrow$3$\alpha$ and $^{16}$O$\rightarrow$4$\alpha$
 reactions. Fig. \ref{fig:2}(a) displays the squared transverse momentum
 distribution of $\alpha$ particles in the laboratory system for the channel
 $^{14}$N$\rightarrow$3$\alpha$+X. The transverse momenta P$_{T}$ are calculated
 by the equation (\ref{eq3}) 
\begin{equation}
{P_{T}=p_{0}\cdot A\cdot sin\theta \label{eq3}}
\end{equation}
 that is, the analysis of P$_{T}$ distributions implies virtually an analysis of
 $\alpha$ particle angular distributions. This distribution has a break at
 P$^2_{T}$=0.05~(GeV/c)$^2$. The continuous line is the sum of two Relay
 distributions.\par
\indent  The values of the momenta in a 3 $\alpha$ particle system can be obtained
 by (subtraction) taking into account the momentum which is acquired by the
 interacting system:
\begin{equation}
   \bf P^{*}_{Ti}\cong {\bf P}_{Ti}- \frac{\sum{\bf P}_{Ti}}{3}\label{eq4} 
 \end{equation}
 \indent The P$_{T}$ distribution for $\alpha$ particles in
 $^{14}$N$\rightarrow$3$\alpha$+X reactions is given in Fig. \ref{fig:2}(b).
 The average P$^*_{T}$ values were expected to be much smaller than the P$_{T}$
 values in the laboratory system and identical for $^{14}$N, $^{12}$C \cite{Belaga95},
 $^{16}$O \cite{Avetyan96} within errors.\par
\indent In order to estimate the energy scale of production of 3$\alpha$ particle
 systems in the $^{14}$N$\rightarrow$3$\alpha$+X channel, we present the invariant
 excitation energy Q distribution  with respect to the $^{12}$C ground state: 
 \begin{equation}
   Q=M^* - M\label{eq5} 
 \end{equation}
 where M is the mass of the ground state corresponding to the charge and the
 weight of the system being analyzed, M$^*$  the invariant mass of the system of
 fragments.
 \begin{figure}
    \includegraphics[width=70mm]{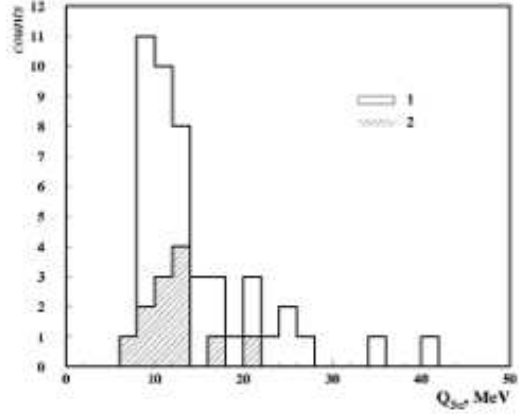}
    \caption{\label{fig:3} The invariant excitation energy Q$_{3\alpha}$
 distribution  of three $\alpha$ particles with respect to the $^{12}$C ground
 state for the process $^{14}$N$\rightarrow$3$\alpha$+X. The following notation
 is used$:$ 1) all the events of the given dissociation, 2) \lq\lq white\rq\rq stars.}
\end{figure}
 \begin{equation}
    {M^{*2}=-(\sum P_j)^2} \nonumber\\
 \end{equation}
  ,P$_{j}$ the 4-momenta of the jfragments.\par
  \indent The main part of the events is concentrated in the Q area from 10 to
 14 MeV, covering the known $^{12}$C levels (Fig. \ref{fig:3}). Softening of the
 conditions of the 3He + H selection, for which the target fragment production is
 allowed, does not result in a shift of the 3$\alpha$ excitation peak. This fact
 suggests the universality of the 3$\alpha$ state population mechanism.\par
 \begin{figure}
    \includegraphics[width=90mm]{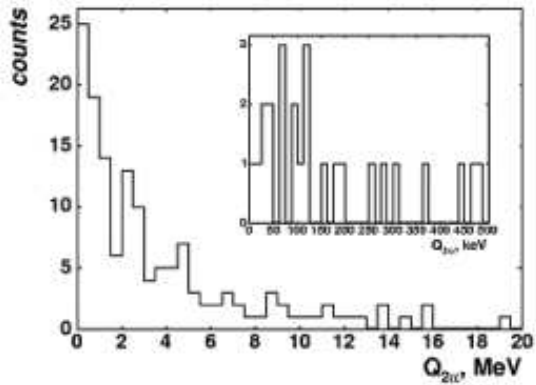}
    \caption{\label{fig:4} The invariant excitation energy Q$_{2\alpha}$ distribution
 of $\alpha$ particle pairs for the process $^{14}$N$\rightarrow$3$\alpha$+X. In
 the inset: a fraction of the distribution at 0-500~keV.}
\end{figure}
 \indent  To estimate the fraction of the events involving  the production of an
 intermediate $^8$Be nucleus in the reactions $^{14}$N$\rightarrow ^8$Be+X$\rightarrow$3$\alpha$+X
 we present the invariant excitation energy distribution for an $\alpha$ particle
 pair with respect to the $^8$Be ground state (Fig. \ref{fig:4}). The first
 distribution peak relates to the value to be expected for the decay products
  of an unstable $^8$Be nucleus in the ground state 0$^+$. This part of the
 spectrum increased by a factor of 20 is presented in the inset of Fig. \ref{fig:4}.      
The distribution centre is seen to coincide well with the decay energy of the
 $^8$Be ground state. The fraction of the $\alpha$ particles originating from the
 $^8$Be decay  is 25-30\%.\par
\begin{figure}
    \includegraphics[width=70mm]{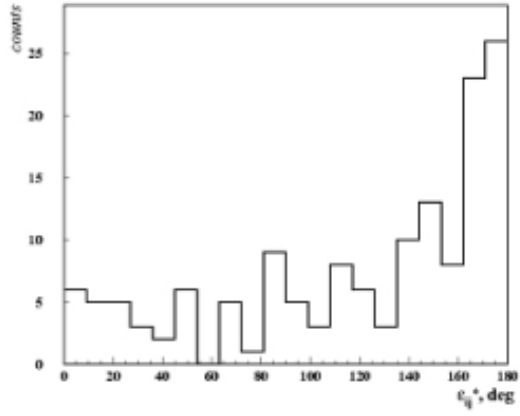}
    \caption{\label{fig:5} The asymuthal angle $\varepsilon_{ij}^*$ distribution in the rest system of 3$\alpha$ particles for the process
$^{14}$N$\rightarrow$3$\alpha$+X.}
\end{figure}

 \indent The role of $^8B$e is clearly pronounced in a strongly asymmetric
 $\varepsilon^*_{ij}$  distribution of $\alpha$ particle pairs in the rest system
 of the 3$\alpha$ partcles of the nucleus $^{14}$N$\rightarrow$3$\alpha$+X 
(Fig. \ref{fig:5}). The asymmetry of the asymuthal angle $\varepsilon^*_{ij}$ is
 also revealed for alpha fragments  in the c.m.s. due to $^{12}$C \cite{Belaga95}
 and $^{16}$O \cite{Avetyan96} decays. The values of the coefficients of asymuthal
 asymmetry and collinearity coincide within errors for $^{14}$N and $^{12}$C and
 they considerably differ the $^{16}$O$\rightarrow$4$\alpha$ decay. This may be
 explained by a more complicated combinatority of $\alpha$ particles for the
 latter nucleus, as well as by the fact that $\alpha$ fragments can also be decay
 products of other intermediate unstable objects.\par   

\section{\label{sec:level6}Conclusion} 
\indent In conclusion we summarize the main results of the present paper. We give
 the results of the study of the dissociation of $^{14}$N nuclei of a primary
 momentum of 2.86~GeV/c per nucleon in their interactions with the emulsion nuclei.
 The charge topology distribution  indicates the leading role of the 2+2+2+1 charge
 configuration channel.\par
\indent  Preliminary data show that for the $^{14}$N$\rightarrow$3$\alpha$+H
 channel the relation of the protons and neutrons N$_p$:N$_d \approx$2, and the
 relation of the helium isotopes $^4$He$:^3$He$\approx$1.5. There are also a few
 helium isotopes (5\% of the total number of interactions) identified as $^6$He.
 A more detailed analysis is needed here.\par
\indent The $\varepsilon^*_{ij}$ distribution in the c.m.s. of $\alpha$ fragments
 for $^{14}$N, $^{12}$C , and $^{16}$O are asymmetric  with an abundance at
 140$^{\circ}$ –- 180$^{\circ}$.\par
\indent The energy scale of the 3$\alpha$ system production has been estimated.
  According to the available statistics 80\% of interactions are concentrated
 at 10-14~MeV. The fraction of the $^{14}$N$\rightarrow ^8$Be+X$\rightarrow$3$\alpha$+X
channel involving the production of an intermediate $^8$Be nucleus is about 25\%.\par 
\begin{acknowledgments}
\indent The work was supported by the Russian Foundation for Basic Research
 ( Grants 96-159623, 02-02-164-12a,03-02-16134, 03-02-17079 and 04-02-16593 ),
 VEGA 1/9036/02.  Grant from the Agency for Science of the Ministry for Education of the
 Slovak Republic and the Slovak Academy of Sciences, and Grants from the JINR
 Plenipotentiaries of the Republic of Bulgaria, the Slovak Republic, the Czech Republic
 and Romania in the years 2002-2005.\par 
\end{acknowledgments}     

\end{document}